\documentclass[twocolumn,showpacs,preprintnumbers,amsmath,amssymb]{revtex4}
\usepackage{amsfonts}
\usepackage{amsmath}
\usepackage{graphicx}
\usepackage{subfigure}
\usepackage{dcolumn}
\usepackage{bm}
\usepackage{overpic}
\usepackage{booktabs}
\usepackage{color}
\usepackage{overpic}
\begin{document}
\title{Hidden space reconstruction inspires link prediction in complex networks}
\author{Hao Liao$^{1,2}$, Mingyang Zhou$^{1,3}$\footnote{zhoumy2010@gmail.com}, Zong-wen Wei$^{1,3}$, Rui Mao$^{1}$,Alexandre Vidmer$^{1,2}$Yi-Cheng Zhang$^{2}$}

\affiliation{1 Guangdong Province Key Laboratory of Popular High Performance Computers, College of Computer Science and Software Engineering, Shenzhen University, Shenzhen 518060, P.R. China\\
2 Department of Physics, University of Fribourg, Chemin du Mus\'{e}e 3, CH-1700 Fribourg, Switzerland\\
3 Department of Modern Physics, University of Science and Technology of China, Hefei 230027, P. R. China.}

\date{\today}
\begin{abstract}

\textbf{As a fundamental challenge in vast disciplines,
link prediction aims to identify potential links in a network based on the incomplete observed information, which has  broad applications ranging from uncovering missing protein-protein interaction to predicting the evolution of networks. One of the most influential methods rely on similarity indices characterized by the common neighbors or its variations.
We construct a hidden space mapping a network into Euclidean space based solely on the connection structures of a network. Compared with real geographical locations of nodes, our reconstructed locations are in conformity with those real ones. The distances between nodes in our hidden space could serve as a novel similarity metric in link prediction. In addition, we hybrid our hidden space method with other state-of-the-art similarity methods which substantially outperforms the existing methods on the prediction accuracy. Hence, our hidden space reconstruction model provides a fresh perspective to understand the network structure, which in particular casts a new light on link prediction.}
\end{abstract}

\maketitle

\section{Introduction}

In recent decades, we have been challenged by understanding the organization of real networks~\cite{newman2003structure,albert2002statistical,WPSGB11}.
Many different disciplines such as Information technology, Biology, Physics, etc, have been studying the organization of real networks~\cite{LNK07}. One of the crucial tasks in complex networks is to reduce the noise and fill up vacant records in large and sparse networks~\cite{Nature1,Nature2}. Vacant records not only refer to the past missing connections between nodes, but also to the future connections. Besides, link information can  detect hidden relationships between terrorists~\cite{Spezzano2014}, predict coverage of a certain virus, identify spurious connections of networks~\cite{An12}, and so on.

The essential problem for link prediction is to measure the
likelihood between nodes accurately~\cite{LJN2010,LSTD10}. A straightforward method to measure similarity is based on the number of common neighbors between two nodes. However this method favors large-degree nodes. In order to overcome the disadvantage of common neighbor method. Some weighted methods are proposed, such as the Jaccard index~\cite{Jaccard1901}, the Salton index~\cite{Salton1975} and the Resource allocation~\cite{RA2009}. These methods based on local information have attracted a great attention due to their efficiency and low computing complexity. Moreover, to suppress imbalance of popular nodes' attractivity and overcome cold start problem~\cite{Wang2016,Leroy2010,ZLZZ09}, various methods based on global information were introduced in link prediction, for example, Simrank~\cite{GJ2002}, hierarchical random graph~\cite{Nature2}, stochastic block model~\cite{BM2011,icdml13}. However, global information based methods are computationally intractable, which limits their application in large complex networks~\cite{LPsurvey,LuZ2010}.

Our main motivation for this paper comes from the networks in which nodes possess a real geographic location such as power grid network. In those networks, the costs restrict the geographic location (i.e., energy cost for power grid network, or efficiency cost for the road network), which shapes the network connections.

Generally, most of real-world networks are lack of real geographic information, and Ref. ~\cite{clustering} suggests that most of these networks populate in some hidden metric space, where the proximity rule governs the connection, that is, the closer nodes are in a hidden space,
the more likely that they are linked together~\cite{space,science13}.
A typical example is the homophily effect in social networks~\cite{mcpherson2001birds,Stai2016A}. Hidden space theory can be used to devise efficient network routing strategy~\cite{n1,n2},
or community detection algorithm~\cite{com}, just name a few.

Here we leverage the proximity rule in hidden space to link prediction by embedding
networks into Euclidean space based on the modified normal matrix~\cite{hlp,hlp1}.
Considering previous hidden space model used in link prediction problem without showing explicit correlation between their model and real existing space. Here, we demonstrate that there is a marked positive correlation for the distance between nodes in hidden space and corresponding real geographical space. Then we predict potential links existing between those pairwise nodes with similar hidden locations.

This paper is organized as follows: In section 2, we first illustrate our hidden space reconstruction process of the network and give a real Italy power grid network example to verify our reconstruction efficiency. Section 3, the application of our hidden space model link prediction is presented. We highlight the achievement of this paper in section 4. Finally in section 5, we introduce theoretical analysis of our hidden space method, and introduce other state-of-the art link prediction methods.

\section{Results}
We start by a brief introduction of the hidden space.
Then we give our hidden space reconstructing process based on advanced normal matrix  in section $2$. Furthermore, distance of nodes in hidden space is utilized to evaluate similarity of non-existing edges in section 3.a. Finally the experimental results are shown in section 3.b and 3.c.

\subsection{Hidden space reconstruction}

Consider an unweighted undirected network $G(V,E)$, where $V$ is the set of nodes and $E$
is the set of links that connect the nodes. There can be only one link between each pair of nodes,
and self-connections are not allowed. The neighbors of a node is the set of nodes that are connected
to it by a link. Link prediction is achieved by calculating a similarity score $s_{ij}$
for each pair of nodes $i$ and $j$ in $V$. This score measures the likelihood for node $i$ and $j$
to be connected with a link. Since $G$ is undirected, the score is symmetric, i.e. $s_{ij} = s_{ji}$.
Then, we sort those nonexsiting links in a descending order by similarity scores.
The scores at the top of the list correspond to links that are the more likely to exist according to
the chosen link prediction method. Therefore how the similarity scores are calculated is the key problem.

The previous methods mainly employ characteristics of neighbor nodes to measure similarity. In this paper, the
hidden space behind the observed network is extracted to characterize the similarity. In some practical networks,
such as power grids, airports, and road networks, nodes usually have fixed locations and connect to geographically closer nodes with higher probability. The probability that a link exists between two nodes is negatively correlated to the distance between the two nodes \cite{Boguna2012Uncovering,Mangioni2013}, such that ${p_{ij}}\propto d_{ij}^{-\beta}$, with $d_{ij}$ the distance of node $i$ and $j$, and $\beta$ a
tunable parameter ($\beta>0$). The fact that nodes preferentially connect to geographically closer nodes is present in the network
with ground truth location (i.e. networks in which nodes possess a fixed location in reality),
and also in some social networks, people excise in a certain areas, but are restricted by financial
and time costs, people living in the same area are more likely to build friendships.Recent empirical experiments reveal that
online social networks also have spatial aggregating characteristics that users in the same region have higher connection
density than across different region, since people in the same region have similar interests and customs \cite{Roick2013Location,Bao2015Recommendations}.

Thus an underlying metric space that determines the topological connection has strong relationship with the geographic location.
Nodes' location can be utilized to measure the similarity of two nodes and predict potential links.
However it is difficult to obtain geographic coordinates for many networks.
Besides real network connections are also influenced by mountains, valleys and rivers, which are not reflected in nodes'
geographic coordinates. Therefore extracting the hidden space is crucial for the understanding of the underlying mechanism of networks.

Though there exists some investigations that apply underlying hidden space to navigation and community detection
(categorization of nodes into groups) \cite{Serrano2012Uncovering,Kleineberg2016Hidden}, the relationship between network structure and the nodes' location in the underlying space, as well as the connection between real and underlying space, is still far from understood.

In this paper, we discuss the network embedded into a $d$-dimension Euclidean space based on the adjacency matrix representing the links between
the $n$ nodes $\mathbf{A}=(a_{ij})_{n\times n}$ of a network, with $a_{ij}=1$ representing the existence of a link between nodes $i$ and $j$, and $a_{ij}=0$ if no link is present. Previous research \cite{Capocci2005Detecting} reveals that similar nodes aggregate together in the spectral space of the Laplacian matrix $L=K-A$ and normal matrix $\mathbf{N}=\mathbf{K}^{-1}\cdot \mathbf{A}$, where $\mathbf{K}=diag\{k_1,k_2,...,k_N\}$ with $k_i$ the degree of node $i$ (the number of links it is connected to). For community detection, the normal matrix usually outperforms
the Laplace matrix \cite{Capocci2005Detecting}, implying that the normal matrix reveals the hidden space better. The maximal eigenvalue
of the normal matrix is 1 (trivial eigenvalue) and corresponds to the eigenvector
$\mathbf{v}_1=(1,1,...,1)_{\mathbf{N}\times 1}^T$. The other $n-1$ non-trivial eigenvalues are in the range $(0,1)$ and the eigenspace
characterized by the non-trivial eigenvectors reflect the topological structure.

Matrix $\mathbf{N}$ could represent the process of heat conduction \cite{Zhou2010}. In heat conduction, each node absorbs heat according to the average temperature of its neighbors. Whereas in practical scenarios, the heat capacity of a nodes may not be linearly proportional to node degree \cite{Zhou2010,wang2006traffic,zhou2012traffic}.
In order to take
this fact into account we introduce a tunable parameter:
\begin{equation}
 \label{eq:norm}
\mathbf{N}_{\alpha}=\mathbf{K}^{-\alpha}\cdot \mathbf{A},
\end{equation}
where $\mathbf{K}^{-\alpha}=diag\{k_1^{-\alpha},k_2^{-\alpha},...,k_n^{-\alpha}\}$. $\mathbf{N}_{\alpha}$ degenerates
into the normal matrix $\mathbf{N}$ when $\alpha=1$.

We use the eigenspace of $\mathbf{N}_\alpha$ to create the hidden space.
Suppose that $\lambda_i$ ($\lambda_1>\lambda_2>...>\lambda_n$) are the eigenvalues of the matrix $\mathbf{N}_{\alpha}$ and that the corresponding eigenvectors
are $\mathbf{v}_1,\mathbf{v}_2,...,\mathbf{v}_n$ ($\|\mathbf{v}_i\|=1$). After removing the trivial eigenvector $\mathbf{v}_1$, given the dimension $d$ of hidden space,
and then construct the hidden space with
\begin{equation}
\label{eq:norm}
\mathbf{W}=Span\{\mathbf{v}_2,\mathbf{v}_3,...,\mathbf{v}_d\},
\end{equation}
where $d>2$ and the span refer to the set of all linear combinations of the elements of $v$. The coordinate $\mathbf{c}_i$ of node $i$ in the hidden space is $\mathbf{c}_i=(v_{2i},v_{3i},...,v_{di})$.
Empirical experiments in many networks suggest that embedding a real network into a small dimension space could reproduce its effective navigation~\cite{Liben2005,Watts2002}. Therefore, we map networks into Euclidean space with dimensions smaller than 40. In section 3.b, experimental results illustrate the effectiveness of our method.

After constructing the hidden space, the distance between node $i$ and $j$ is $d_{ij}=||\mathbf{c}_i-\mathbf{c}_j||$.
Since nodes prefer to connecting geographically nearby nodes, we take the negative value of the
distance as the similarity score, $s_{ij}=-d_{ij}$. Non-existing links with top-$d$ scores are predicted as potential links.

\subsection{Correlation between hidden space and real space}

\begin{figure*}
\centering
\includegraphics[width=6.5in]{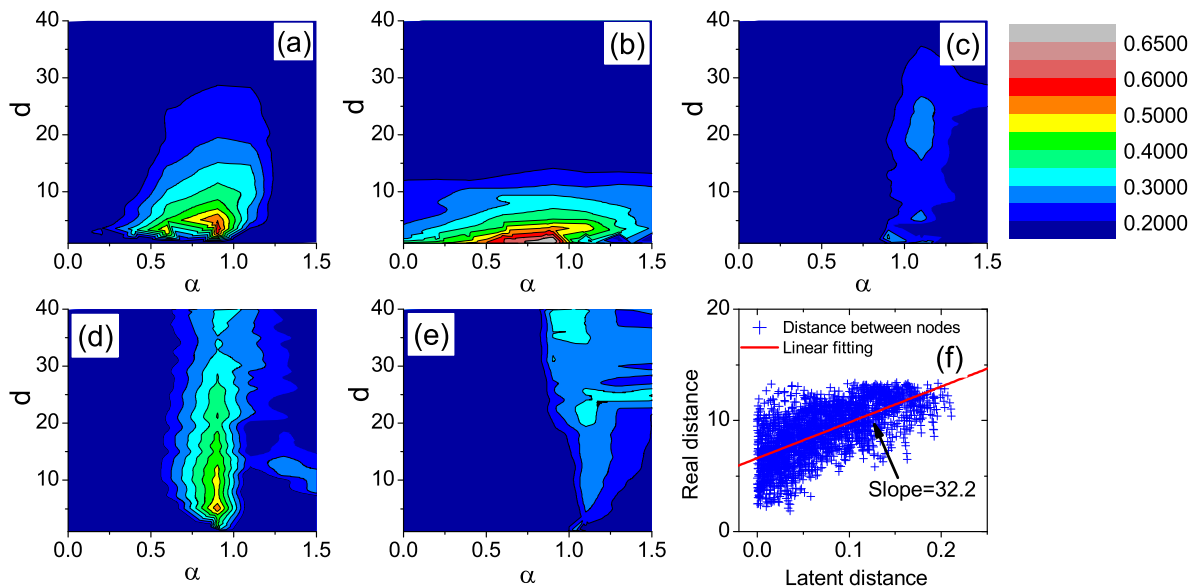}
  \caption{(Color online)
Panels $(a)\sim(e)$ show the \textbf{Spearman correlation} coefficients as a function of $\alpha$ and dimension $d$ of hidden space.
In panel (a), an artificial network is generated by Newtonian model with size $N=700$,
average degree $<k>=4$,$\gamma=2.5$ and $\beta=2$.
Results are the simulation of 100 independent experiments.
(a) Model network. (b) Italian PowerGrid network. (c) Maayan-faa network. (d) OpenFlights network. (e) Euroroad network. Panel (f) shows the scatter plot of real distance versus distance in the hidden space for $\alpha=0.85$ and $d=2$.
}
\label{fig:spearman}
\end{figure*}

\begin{figure}
\center
\begin{overpic}[scale=0.7]{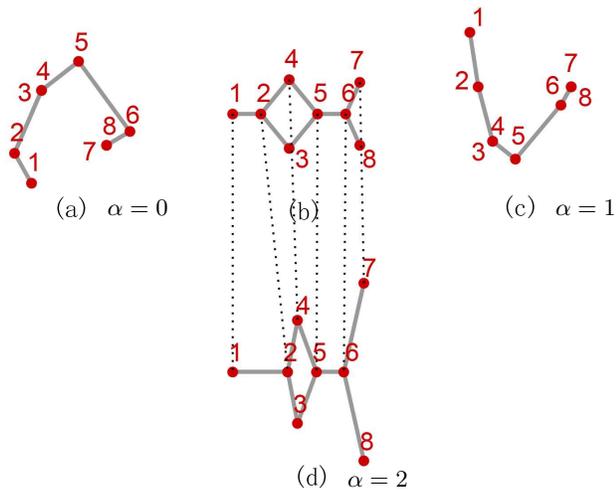}
\label{fig:modelcorrelation}
    \put(18.5,43.5){\small \color{black}{$\alpha=0$}}
    \put(87,43.5){\small \color{black}{$\alpha=1$}}
    \put(55.2,2){\small \color{black}{$\alpha=2$}}
\end{overpic}
\caption{Skeleton illustration of the extraction of the hidden space in a model network. All the four subfigures are
plotted according to their geographic or hidden position. (a) Hidden location with $\alpha=0$.
(b) Real geographic location.
(c) Hidden location with $\alpha=1$.
(d) Hidden location with $\alpha=2$.}

\end{figure}

We explore the relation between the hidden space coordinates and the real geographical locations. We find the distance between the nodes in the real space is strongly correlated with those of the
hidden space. In Figure \ref{fig:spearman}, we show the Spearman correlation between hidden and real
geographical locations in real networks as a function of $\alpha$ and dimension $d$. For most networks, the maximal correlation is around $\alpha=1$. Table \ref{tab:optimalparameter} shows the maximal pearman correlation and the corresponding optimal $\alpha$ and dimension $d$ for different networks. For Italian PowerGrid network, the optimal dimension $d=1$ due to the linear outline of Italian Map (See Fig. \ref{fig:modelcorrelation}). The optimal dimension of the model network is $d=3$, which is different from real geographical dimension $d=2$. It is because that apart from location factors, degree distribution also shape the network, which is reflected in the additional one dimension.
Euroroad network has a much larger dimension $d_{optimal}=6$ than other networks, meaning that Euroroad structure is determined by many non-geographical underlying factors such as policy, country, economic and so on. In Euroroad network, when $d=3$ and $\alpha=1$, $Spearman=0.4815$ is close to the optimal $Spearman_{optimal}=0.5047$, implying that geographical location dominates the main body of Euroroad network.

\begin{figure*}
\center\scalebox{1}[1]{\rotatebox{0}{\includegraphics[height=4in]{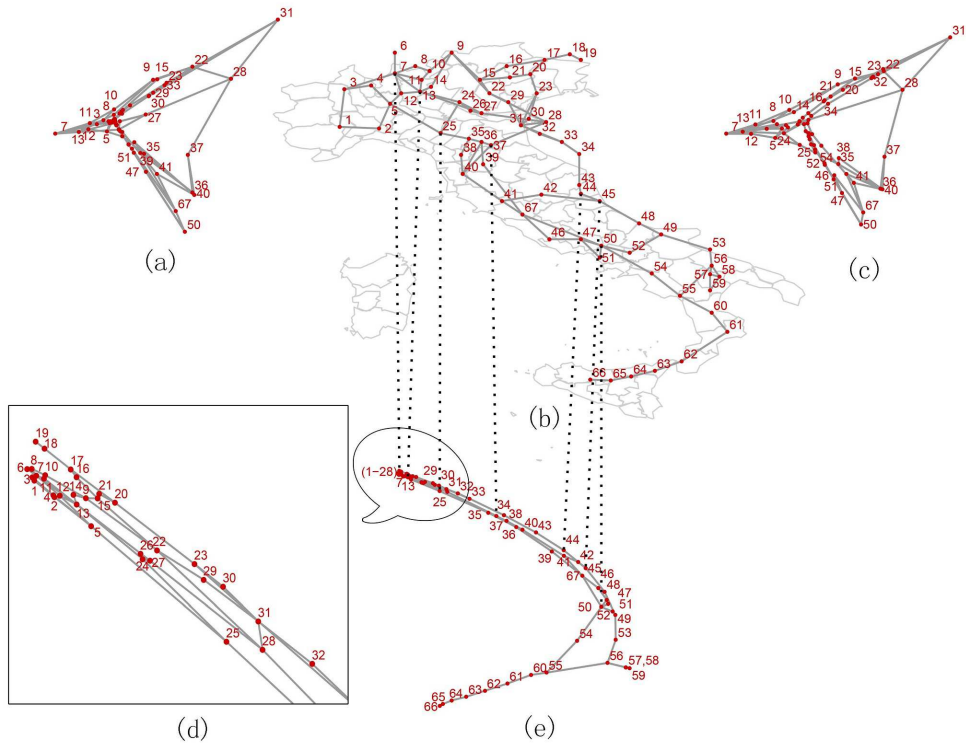}}}
\caption{Real geographic position and hidden location as a function of $\alpha$ for Italy Power-Grid network.
All the four subfigures are plotted according to their geographic or hidden position. (a) Hidden location with $\alpha=-1$.
(b) Real geographic location. (c) Hidden location with $\alpha=0$. (d) Hidden location with $\alpha=1$.}\label{fig:italycorrelation}
\end{figure*}

To better understand the relationship between the hidden and real geographic locations, we further compare different value $\alpha$ in Figure \ref{fig:modelcorrelation} and Figure \ref{fig:italycorrelation}.
In Fig. \ref{fig:modelcorrelation}, the hidden locations reveal the real geographic location in the skeleton illustration network with optimal value of $\alpha=2$.

In the following paper, as we are interested in the hidden relationship between nodes, in real networks, we take into account the value that maximizes the correlation
between the hidden distances and real distances of all pairwise nodes.

\begin{table}
\caption{Maximal Spearman correlation and the corresponding optimal $\alpha$ and optimal dimension $d$ for different networks.}
\label{tab:optimalparameter}
\begin{center}
\begin{tabular}{p{2.6cm} p{2cm} p{1.5cm} p{1.5cm}}
\hline
\hline
Network &$Spearman$ &$\alpha_{optimal}$ &$d_{optimal}$ \\
\hline
Model network        &0.6235 &0.85 &3\\
Italian PowerGrid           &0.6727 &0.85 &1\\
Maayan-faa               &0.3592 &0.95 &2\\
OpenFlights              &0.5926 &0.95 &2\\
Euroroad              &0.5047&1&6\\
\hline
\hline
\end{tabular}
\end{center}
\end{table}

\section{Link prediction in real-world networks}

\begin{figure}
\center\scalebox{0.6}[0.6]{\rotatebox{0}{\includegraphics[width=5in]{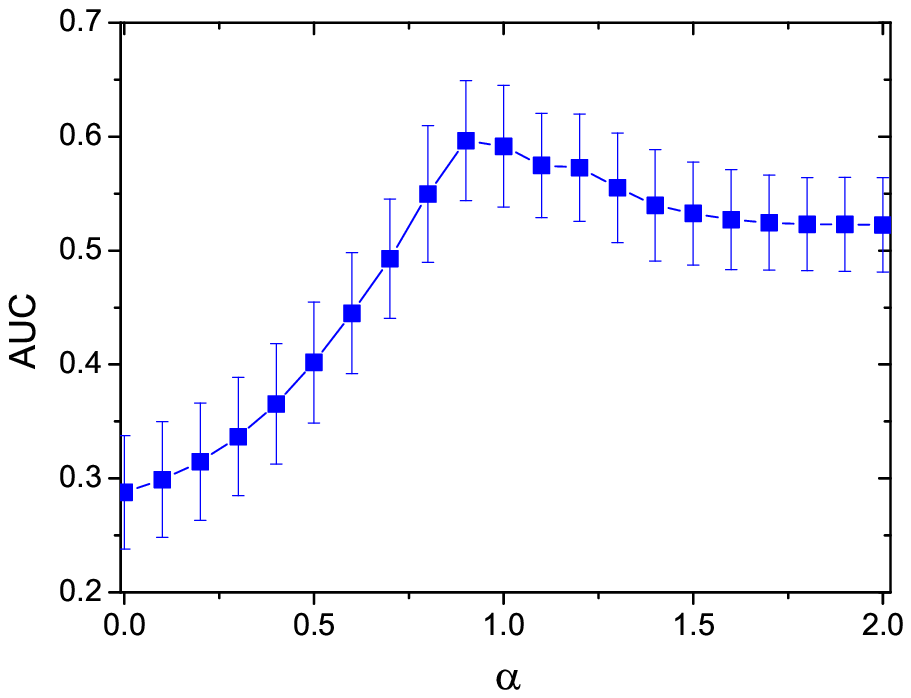}}}
\caption{AUC of model networks as a function of $\alpha$. The results are the average of 10 independent simulations. The optimal AUC by equation \ref{eq:theoryauc} is 0.57 that matches the experimental simulations. }\label{fig:lustrationmodel}
\end{figure}

\subsection{Coordinate determination}

In link prediction, the set of observed links $M$ of a network is randomly divided into two parts: the training set $M^T$ treated
as known information, and the probe set $M^p$, used to verify the accuracy of the prediction.
The information contained in the probe set is considered as unknown and is not used during the prediction process.
The addition of the two set, $M^T$ plus $M^p$, is equal the whole data set.
Besides, disconnected nodes in the training set are not considered. We choose the training set to contain $90\%$ of the links and the probe set $10\%$ of the links. The aim of link prediction is to use links in training set to predict probe set as accuracy as possible.

Note that, only training set is used to reconstruct the hidden space underlying a network in our experiments. Based on the hidden locations of all nodes, links between pairwise nodes with close locations are predicted as potential links. Each link $e_{ij}$ is assigned a score $s_{ij}=-d_{ij}$. Links ranking in the top-$L$ list are predicted as potential links in probe set.

In this paper, we employ a standard metric, area under the receiver operating characteristic curve (AUC)~\cite{AUC1984} to measure
the accuracy of the prediction. AUC can be interpreted as the probability that a randomly chosen missing link from $M^p$
is given a higher score than a randomly chosen nonexistent link.
Then, AUC requires $n$ times of independent comparisons. we randomly choose a missing
link and nonexistent link to compare their scores. After the comparison, we record there are
$n_1$ times the missing link having a higher score, and $n_2$ times they have the same score.
The final AUC is calculated as $AUC=(n_1+0.5\times{n_2})/n$. If all the scores are given by
an independent and identical distribution, then $AUC$ should be around $0.5$. A higher AUC
is corresponding to a more accurate prediction.

The key issue of the proposed method is to determine optimal parameter $\alpha$ and $d$.
For a given network with geographical location, optimal $\alpha$ and $d$ could be obtained by comparing hidden space and real location (See Fig. \ref{fig:spearman}). For many networks without geographical location, according to empirical studies in Table \ref{tab:optimalparameter}, we firstly fix $d=3$ and calculate the $AUC$ as a function of $\alpha$, from which we could obtain local optimal $\alpha_{optimal}$. Then, we set $\alpha=\alpha_{optimal}$ and calculate the $AUC$ as a function of $d$, from which local optimal $d_{optimal}$ is obtained. Experiments in real networks reveal that optimal $\alpha_{optimal}$ is around 0.95, and the dimension of hidden space is less than 10 (See Table \ref{tab:compareauc}). (We could also firstly fix $\alpha$, and later fix $d$. However according to Fig. \ref{fig:spearman}, Spearman correlation have few fluctuation at optimal $d$, whereas it changes sharply at optimal $\alpha$. Therefore it is better to set $d$ first, and later consider $\alpha$.)

\subsection{Empirical analysis}
We apply the hidden space method to six real networks, which all exist in the physical world. The first four networks
possess real locations, The last two are protein-protein interaction network without physical locations. All the simulations in this section are the average of 50 different divisions of the dataset.

(1)PowerGrid \cite{konect}: the electrical power grid of western US, with nodes representing generators, transformers and substations,
and links corresponding to the high-voltage transmission lines between nodes. This network contains 4,941 nodes,
and they are well connected.

(2) Maayan-faa \cite{konect}: This networks represent the flight routes in the USA. The nodes are airports and
the links represent the presence of a flight route between two airports.
This is an directed and unweighted network, containing 1,226 nodes and 2,615 edges.

(3) OpenFlights \cite{konect}: a directed network containing flights between airports of the world, in which directed edge represents a flight from one airport to another.  Here it has 2939 nodes and 30501 edges.

(4) Euroroad \cite{konect,eroads}: An undirected and unweighted network representing the international roads connecting the cities in Europe (E-roads).
The nodes represent the cities and the links represent the roads. The network contains 1,174 nodes and 1,419 edges.


(5) Yelp  \cite{Yelp}: An undirected and unweighted social network in the round 4 of the Yelp
academic challenge dataset.
Yelp is a website where users can review and rate various
businesses such as restaurants, doctors, and bars. For our analysis, we keep only the users who have at least one
friend. The network contain 123368 nodes and 1911997 edges. In the paper, we sample 1417 connected nodes randomly with 4472 edges and keep their connections.

(6) Maayan-pdzbase \cite{konect,pdzbase}: a network of protein-protein interactions, which is an undirected and unweighted network, containing 212 nodes and 244 edges.
We only take into account of the giant connected component of these networks. This is because for a pair of nodes located in two disconnected components, their similarity score will be zero according to most prediction methods. Moreover self-loop links and nodes' direction are ignored for convenience. After these data processing, Table 1 shows the basic statistics of all the giant components of those networks.


\begin{table*}
\caption{Structural properties of the different real networks. Structural properties include network size ($N$), edge number ($E$), degree Heterogeneity ($H=\langle k^2\rangle/\langle k\rangle^2$), degree assortativity ($r$), clustering coefficient ($\langle C \rangle$) and average shortest path length ($\langle d \rangle$). }
\label{tab1}
\begin{center}
\begin{tabular}{p{2.6cm} p{1cm} p{1cm} p{1cm} p{1cm} p{1cm} p{1cm} p{1cm} p{1cm} p{1cm}}
\hline
\hline
Network &$N$ &$E$ &$H$ &$r$ &$\langle C \rangle$ &$\langle d \rangle$ \\
\hline
PowerGrid           &4941 &6594 &1.450 &0.004 &0.015  &18.989\\
Maayan-faa               &1226 &2408 &1.873 &-0.015 &0.012 &5.929 \\
maayan-pdzbase          &161 &209 &2.263 &-0.466 &0.001 &5.326\\
Yelp                   &450 &940 &2.163 &-0.023 &0.045  &4.188\\
OpenFlights              &2905 &15645 &5.184 &0.049 &0.054 &4.097\\
Euroroad              &1039 &1305 &1.229 &0.090 &0.005 &18.395\\
\hline
\hline
\end{tabular}
\vspace*{0.0cm}
\end{center}
\end{table*}

\begin{figure*}
\centering
\includegraphics[width=5in,height=2.2in]{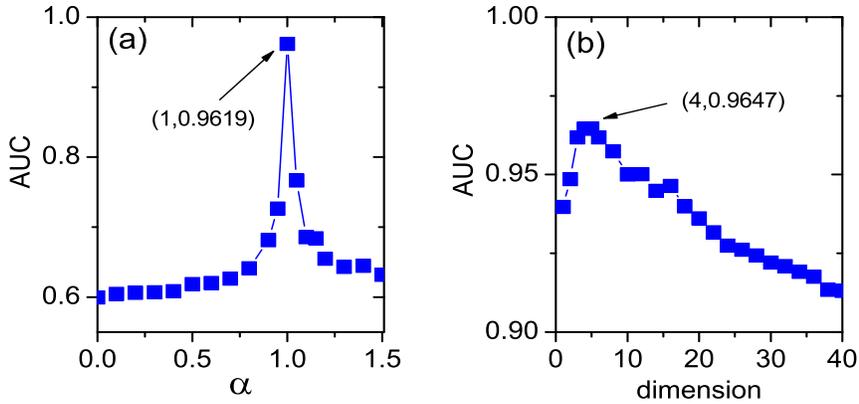}
\caption{The approach to obtain optimal $\alpha$ and $d$ in US PowerGrid network. (a) $AUC$ as a function of $\alpha$ at $d=3$. (b) $AUC$ as a function of dimension $d$ at $\alpha=1$.}
\label{fig:auc}
\end{figure*}

Figure.~\ref{fig:auc} shows the local optimal $\alpha$ and $d$ in US PowerGrid network. Fig.~\ref{fig:auc}(a) plot $AUC$ as a function of $\alpha$ when $d=3$, and $\alpha_{optimal}=1$. Note that optimal $\alpha_{optimal}$ of most network is smaller than 1 (refer to Table \ref{tab:compareauc}). $AUC$ varies sharply around 1. Figure.~\ref{fig:auc}(b) shows $AUC$ as a function of dimension $d$ where $d_{optimal}=4$, and $\alpha_{optimal}=1$. Besides, $AUC$ becomes stable around $d_{optimal}=4$. When the dimension is $d=[2,6]$, $AUC$ fluctuates within 2\%, revealing great robust to $d$. Similar to US PowerGrid network, we also obtain the optimal $\alpha$ and $d$ for other network in Table \ref{tab:compareauc}. Notice that though $\alpha_{optimal}$ is around 1, we cannot determine $\alpha_{optimal}=1$, since $AUC$ will decrease sharply from $\alpha_{optimal}$ to 1.

%

Further, comparing our method, the hidden space method (HS), with five stat-of-art similarity indices, \emph{Common Neighbor} (CN),
\emph{Jaccard coefficient} (Jaccard) \cite{Jaccard1901}, \emph{Resource allocation} (RA)~\cite{RA2009},
Katz ~\cite{Katz1953}, and Structural Perturbation Method (SPM) \cite{lu2015toward} . In these indexes,
two nodes are considered to be similar if they have common important topological features~\cite{LuZ2010}.
The results are shown in Table~\ref{tab:compareauc}.
CN, AA, RA are local index methods, while Katz and SPM are global information based methods. The AUC of hidden space method
is substantially higher than all local indexes on most networks, also better than SPM method. Katz index performs better than our method in three networks but costs extremely high computation complexity.

\begin{table*}
\caption{AUC comparison for different methods and different networks. For the hidden space method (HS), we choose
the $\alpha$ value that maximizes AUC. The highest value for each network is indicated in boldface.}
\label{tab:compareauc}
\begin{center}
\begin{tabular}{p{2.6cm} p{1cm} p{1cm} p{1cm} p{1cm} p{1cm} p{1cm} p{1cm} p{1cm} p{1cm}}
\hline
\hline
Network &$CN$ &$AA$ &$RA$ &$Katz$ &$SPM$ &$HS$ &$\alpha$&$d$\\
\hline
US PowerGrid                &0.628 &0.628 &0.841 &\textbf{0.976} &0.595  &0.965&1&4\\
Maayan-faa               &0.621 &0.625 &0.620 &\textbf{0.844} &0.664 &0.840&0.95&6 \\
maayan-pdzbase           &0.481 &0.484 &0.478 &\textbf{0.838} &0.662 &0.751&0.8&6 \\
Yelp                     &0.744 &0.748 &0.750 &0.757 &0.710 &\textbf{0.789}& 0.95  &9\\
OpenFlights              &0.970 &0.973 &0.972 &\textbf{0.980} &0.917  &0.920& 0.97 &6\\
Euroroad                 &0.539 &0.535 &0.538 &0.898 &0.541 &\textbf{0.913}& 0.99&6\\
\hline
\hline
\end{tabular}
\vspace*{0.0cm}
\end{center}
\end{table*}

The performance of link prediction can also be evaluated by precision metric. Precision is the ratio of right predicted links, given a set of potential links. That is to say, if we choose that the $L$ links with the highest scores are the predicted ones,
and $L_r$ links are in the probe set $M^p$, then the precision $P(L)=L_r/L$. Clearly, higher precision means higher
 accuracy. In our experiments, we choose the length of prediction list equal to the size of probe set $L=|M^p|$. Therefore $P(L)\in(0,1)$.

The results for precision are shown in Table~\ref{tab:precision}. We choose the same values for $\alpha$ that in Table~\ref{tab:compareauc}.
HS method achieves high AUC, yet with smaller precision compared to the other methods.

Intuitively, higher accuracy means higher AUC and higher precision. The reason of deviation between AUC and precision is that AUC evaluates
the whole score difference of probe links and non-existing links. Whereas precision only concerns top-$L$ high-score links. Besides, this result still holds under different $L$. Therefore we only present the results at $L = |M^p|$.


\begin{table*}
\caption{Precision comparison for different methods and all the networks. The best performing method for each network is indicated in boldface.}
\label{tab:precision}
\begin{center}
\begin{tabular}{p{2.6cm} p{1.1cm} p{1.1cm} p{1.1cm} p{1.1cm} p{1.1cm} p{1.1cm} p{1.1cm} p{1.1cm} p{1.1cm}}
\hline
\hline
Network &$CN$ &$AA$ &$RA$ &$Katz$ &$SPM$ &$HS$ \\
\hline
US PowerGrid                &0.0126 &0.0241 &0.0220 &\textbf{0.0588} &0.0363  &0.0081\\
Maayan-faa               &0.0226 &0.0141 &0.0098 &\textbf{0.0413} &0.0296 &0.0025 \\
maayan-pdzbase           &0.0000 &\textbf{0.0062} &0 &0.0000 &0.0000 &0.0000\\
Yelp                     &0.0486 &0.0825 &0.0418 &\textbf{0.0828} &0.0500 &0.0251\\
OpenFlights              &0.2101 &0.2361 &0.2571 &0.0012 &\textbf{0.2900}  &0.0038\\
Euroroad                 &0.0055 &0.0026 &0.0019 &\textbf{0.0099} &0.0031 &0.0021\\
\hline
\hline
\end{tabular}
\vspace*{0.0cm}
\end{center}
\end{table*}


\subsection{Hybrid prediction method}
In order to improve the precision of our method, we compare the overlap of prediction list between each other. It is less than 1\% common links between Katz and HS, and around 5\% between CN and HS.
The small overlap links reveals other methods and HS method tend to predict different kinds of potential links. Therefore, they could complement each other's advantages to improve the precision metrics.

Due to the differences between other methods and HS method, we propose a
hybrid approach to enhance the prediction precision. Combining $HS$ with other methods,
we multiply the similarity obtained by HS and those similarity by another methods.
For example, if the similarity score of CN and HS methods for nodes $i$ and $j$ are $s^{CN}_{ij}$ and $s^{HS}_{ij}$,
the hybridized score between node $i$ and $j$ becomes $s_{ij}'=s^{CN}_{ij}\ast s^{HS}_{ij}$. The prediction list
is obtained again by choosing links with top-$L$ score $s_{ij}'$.
We show the results of this hybridization in Tab.~\ref{tab:precision2}. The precision is remarkably improved in most of networks.

\begin{table*}
\caption{Precision comparison for the hybridized similarity scores for all the networks.
Boldface Indicates that hybrid similarities enhance precision than the original methods.}
\label{tab:precision2}
\begin{center}
\begin{tabular}{p{2.6cm} p{1.4cm} p{1.4cm} p{1.4cm} p{1.4cm} p{1.4cm} p{1.4cm} p{1.4cm} p{1.4cm}}
\hline
\hline
Network &$CN'$ &$AA'$ &$RA'$ &$Katz'$ &$SPM'$\\
\hline
US PowerGrid                &\textbf{0.0335} &\textbf{0.0431} &\textbf{0.0321} & 0.0509 &0.0072\\
Maayan-faa               &\textbf{0.0295} &\textbf{0.0275} &\textbf{0.0171} &0.0296 &\textbf{0.0213}\\
maayan-pdzbase           &0.0000 &\textbf{0.0072} &0.0000 &0.0000 &0\\
Yelp                     &\textbf{0.0802} &\textbf{0.0828} &\textbf{0.0603} & 0.0828 &0.0491\\
OpenFlights              &\textbf{0.2277} &0.2368 &0.2570 &0.0012 &\textbf{0.2914}\\
Euroroad                 &0.0014 &0.0007 &0.0014 &\textbf{0.0064} &0.0031\\
\hline
\hline
\end{tabular}
\vspace*{0.0cm}
\end{center}
\end{table*}

\section{Conclusion}
We conclude that network topology and the real location of nodes is strongly affected
by the distance between nodes in the hidden space. Our experimental results on both artificial and real-world networks show that the hidden space locations which are highly correlated with the geographic locations, can be reconstructed merely from connectivity matrix without the knowledge of the real geographic locations. This is a very strong point, as the geographic coordinates are not always available in networks. For instance, we possess only the connections between power stations, and we want to retrieve the distance between them.

In this paper, the hidden space distance are used to predict missing links, giving high similarity scores between pairwise nodes which are geographically close in the hidden space. Our results show that the hidden space method improves AUC significantly.
Additionally, we find an interesting phenomenon that hidden space method obtains high AUC, but low precision.  It means the HS method could find some missing links which cannot be identified by other methods. Since the results on the two metrics are so different for the hidden space method, we complemented it with other methods which significantly enhance the predicting precision.

We believe that the present and future
work on the hidden space and link prediction will deepen our
understanding of the fundamental relationships between
structure and function of complex networks.

\section{Materials and methods}
\subsection{Illustration ground truth location of the network}

To verify the effectiveness of hidden space method, we explore the hidden metric space in an artificial model
network and in the Italian Power-Grid network, which is a network with ground truth location. The Italian Power-Grid is the topology
of Italian high-voltage electrical network, which contains 98 nodes and 175 edges.
Since real networks usually follow scale-free or similar degree distribution, a Newtonian model \cite{Boguna2012Uncovering} is utilized to
generate scale-free networks embedded in metric spaces as follows: Firstly we set the final network size $N$,
then we assign geographic coordinates to each node in the metric space, as well as their expected degree.
Nodes are distributed in a D-dimension space with uniform density and their degree values are generated according
to a power-law distribution $ p_0(k)=c_0k^{-\gamma} $, $ k\in \left[k_0, +\infty\right)$,
where $k_0$ is the minimum expected degree and $c_0$ is a normalization constant. A pair of nodes $i$ and $j$ is connected by an edge
with probability $r(d_{ij},k_i,k_j)=\frac{1}{(1+\frac{d_{ij}}{\mu k_ik_j})^\beta}$. In our experiments,
nodes are distributed in a 2-dimension space $\{x,y\;|\;0\leq x,y\leq 1\}$,
and we set $n=700$, $\gamma=2.5$, $k_0=1$, $\mu=\frac{\beta-1}{2<k>}$ and $\beta=2$ \cite{Boguna2012Uncovering}. The isolated nodes
are removed in the model network.




Based on ground truth location of the model network, theoretical AUC is calculated.
For two random nodes $i(x_i,y_i)$ and $j(x_j,y_j)$ in the 2-dimension square space,
the probability $||\mathbf{d}_i-\mathbf{d}_j||=r$ that the distance between the nodes is equal to a value $r$:
\begin{equation}
 \label{eq:distance}
  \begin{aligned}
p_2(r_{ij}=r)&=\int_0^{\min(r,1)}p_1(l_1)p_1(\sqrt{r^2-l_1^2})\mathrm{d}l_1,\quad 0\leq r\leq \sqrt{2},
\end{aligned}
\end{equation}
where $p_1(|x_i-x_j|=l)=2(1-l),0\leq l\leq 1$ is the coordinate difference probability.

Given a random edge $e_{ij}$ ($e_{ij}=1,0$), the distance of two endpoint nodes has the conditional probability $p_3(r_{e_{ij}}|e_{ij})$,
\begin{equation}
 \label{eq:distanceedge}
\begin{aligned}
&p_3(r_{e_{ij}}|e_{ij})=\int\int_{k_i,k_j}p(r_{e_{ij}},k_i,k_j|e_{ij})dk_idk_j\\
&=\int\int_{k_i,k_j}p(e_{ij}|r_{e_{ij}},k_i,k_j)\frac{p(r_{e_{ij}},k_i,k_j)}{p(r_{e_{ij}})}dk_idk_j\\
&=\begin{cases}
\int\int_{k_i,k_j}r(r_{e_{ij}},k_i,k_j){p_2(r_{e_{ij}})p_0'(k_i)p_0'(k_j)}dk_idk_j,\\
 \qquad\qquad\qquad\qquad\qquad\qquad\qquad\qquad\qquad\text{if } e_{ij}=1,& \\
\int\int_{k_i,k_j}(1-r(r_{e_{ij}},k_i,k_j)p_2(r_{e_{ij}})p_0'(k_i)p_0'(k_j)dk_idk_j, \\
\qquad\qquad\qquad\qquad\qquad\qquad\qquad\qquad\qquad\text{if } e_{ij}=0,& \\
\end{cases}
\end{aligned}
\end{equation}
where $p_0'(k)=\frac{k}{<k>}p_0(k)$ indicates the probability that one endpoint of a random edge has degree $k$ and $<k>$ is the average degree of the network.

Theoretical AUC would be obtained by comparing scores of an existing edge and an non-existing edge, suppose the two edges' score are $s_1=-r_1$ and $s_2=-r_2$ respectively,
\begin{equation}
 \label{eq:theoryauc}
\begin{aligned}
AUC&=\int ds_1\int p(s_1>s_2)\mathrm{d}s_2\\
&=\int p_3(r_{1}|e_{1}=1)\mathrm{d}r_1\int_{r_2\geq r_1}p_3(r_{1}|e_{2}=0)\mathrm{d}r_2.
\end{aligned}
\end{equation}

Figure \ref{fig:lustrationmodel} shows AUC result in the model network. As $\alpha$ increases from 0 to 2, AUC increases sharply in the beginning,
then decreases slowly at $\alpha>1$. Integrating Fig. \ref{fig:spearman} and \ref{fig:lustrationmodel}, the optimal
AUC appears at $\alpha\approx 1$, as expected from the previous results on real geographic location.

\subsection{Indices for link prediction}

(i) \emph{Common Neighbor} (CN).
The idea of this metric is that the more neighbors two nodes $i$ and $j$ have in common,
the more likely they are to form a link. Let $\Gamma(i)$ denote the set of neighbors of node $i$,
the simplest measure of the neighborhood overlap can be the directly calculated as:
\begin{equation}
  \label{eq:cn}
  s^{\rm CN}_{ij} = |\Gamma(i)\cap\Gamma(j)|.
\end{equation}
CN is the method used by most websites [...]. However, the drawback of CN is that it favors
the nodes with large degree. Using the adjacency matrix, $(A^2)_{ij}$ is the number of
different paths with length two connecting $i$ and $j$. So we can rewrite $s_{ij} =(A^2)_{ij}$.
Newman \cite{LuZ2010} used this quantity in the study of collaboration networks, showing the
correlation between the number of common neighbors and the probability that two scientists
will collaborate in the future. Therefore, we here select CN as the representative of all
CN-based measures.

(ii) \emph{Jaccard coefficient} (Jaccard) \cite{Jaccard1901}.
This index was proposed by Jaccard over a hundred years ago. The algorithm is a traditional similarity measurement in the literature. It is defined as
\begin{equation}
  \label{eq:jaccard}
  s^{\rm Jaccard}_{ij} = \frac{|\Gamma(i)\cap\Gamma(j)|}{|\Gamma(i)\cup\Gamma(j)|}.
\end{equation}
The motivation of this index is that the raw number of common neighbors favors the large degree
nodes, simply because the large degree nodes have more neighbors than smaller ones.
The normalization gives more credit to nodes sharing high number of neighbors compared
to their total joint number of neighbors, eventually removing the bias towards high degree nodes.
Note that there are many other ways to remove the tendency of CN to large degree nodes,
such as cosine similarity, Sorensen index, Hub promoted index and so on, see ~\cite{LuZ2010}.

(iii) \emph{Resource allocation} (RA)~\cite{RA2009}.
This index is inspired by the resource allocation dynamics on complex network.
Consider a pair of nodes, $i$ and $j$, which are not directly connected.
Suppose that the node $i$ needs to give some resource to $j$, using common neighbors
as transmitters. Each transmitter (common neighbor) starts with a single unit of resource, and then distributes it
equally among all its neighbors. The similarity between $i$ and $j$ can be the directly calculated as the amount of resource received from
their common neighbors:
\begin{equation}
 \label{eq:RA}
 s^{\rm RA}_{ij} = \sum_{z\in\Gamma(i)\cap\Gamma(j)}\frac{1}{k_{z}}.
 \end{equation}
This measure is symmetric. By using $log(k_z)$ instead of $k_{z}$ in Eq.~\ref{eq:RA},
the index becomes the Adamic-Adar (AA) Index \cite{AA03}.
The difference between RA and AA is small if $k_z$ is small.
However, in heterogeneous networks $k_z$ can be very large, then the differences of RA and AA becomes large.
By giving less contribution to the high degree nodes, RA usually achieves a higher link prediction accuracy than AA.

(iv) Katz ~\cite{Katz1953}.
This index takes all paths between the two nodes $i$ and $j$ into consideration. It is defined as
\begin{equation}
  \label{eq:katz}
  s^{\rm Katz}_{ij}=\alpha A_{ij}+\alpha^2 A_{ij}^2+\alpha^3 A_{ij}^3+\dots,
\end{equation}
where $\alpha$ is a free parameter and $A$ is the adjacency matrix of the network. If the
parameter is small, the index is close to CN. In order for the sum to converge, $\alpha$
must be chosen such that $\alpha<\frac{1}{\lambda_{max}}$. where $\lambda_{max}$ is the maximum
of the eigenvalues of matrix $A$.
When $\alpha<\frac{1}{\lambda_{max}}$, $S^{\rm Katz}_{ij}$ could be simplified as
\begin{equation}
  \label{eq:katz}
  S=(I-\alpha A)^{-1}-I,
\end{equation}
where S=$(s_{ij})_{n\times n}$ and $\lambda_{max}$ is the maximum eigenvalue of adjacent matrix $A$.

(v) Structural Perturbation Method (SPM) \cite{lu2015toward}. This index is based on the hypothesis that the features of a
network are stable if a small fraction of edges is randomly removed. In SPM, we perturb a
network by removing $\Delta E$ edges. The corresponding matrix corresponding to the randomly removed edges is $\Delta A$, the remaining
edges are represented by the matrix $A^R$, with $A=A^R+\Delta A$.
Assume that the perturbation of the eigenvectors of $A$ and $A^R$ is only minor, then the perturbated matrix writes
\begin{equation}
  \label{eq:SPM}
\tilde{A}=\sum _{k=1}^N (\lambda_k+\Delta \lambda_k)x_kx_k^T,
\end{equation}
where $\lambda_k$ and $x_k$ are the eigenvalue and the corresponding orthogonal and normalized eigenvector for $A^R$, respectively,
and $\Delta\lambda_k\approx \frac{x_k^T\Delta Ax_k}{x_k^Tx_k}$. The similarity of nodes $i$ and $j$ is given by the
corresponding value of the matrix
$\tilde{A}$, $\tilde{a}_{ij}$.
\section{ACKNOWLEDGMENTS}
We thank Prof. Matus Medo, Prof. Chi Ho Yeung, Prof. Bing-Hong Wang for fruitful discussion and comments. This work is sponsored by the National Natural Science Foundation of China (Grant No. 11547040), Guangdong Province Natural Science Foundation (Grant No. 2016A030310051, 2015KONCX143), Shenzhen Fundamental Research Foundation (JCYJ20150625101524056, JCYJ20160520162743717, JCYJ20150529164656096), National High Technology Joint Research Program of China (Grant No.2015AA015305), Project SZU R/D Fund (Grant No. 2016047), CCF-Tencent (Grant No. AGR20160201), Natural Science Foundation of SZU (Grant No. 2016-24).

%

\end{document}